\documentclass[prc,twocolumn,superscriptaddress,superscriptfootnote,nofootinbib]{revtex4}
\usepackage{amssymb}

\usepackage{amsmath,amsfonts,amssymb,epsfig,color,MnSymbol}

\begin{document}
\title{Extended multiconfigurational dynamical symmetry}
\author{H. G. Ganev}
\affiliation{Joint Institute for Nuclear Research, Dubna, Russia}
\affiliation{Institute of Mechanics, Bulgarian Academy of Sciences,
Sofia, Bulgaria}

\setcounter{MaxMatrixCols}{10}

\begin{abstract}
An extended multiconfigurational dynamical symmetry (EMUSY) within
the symplectic symmetry approach to clustering (SSAC) is proposed
for the general case of multicluster nuclear systems. A
characteristic property of the EMUSY is that it includes more
general symplectic, i.e. number non-preserving, transformations
which contain the standard number-preserving (unitary)
multiconfigurational dynamical symmetry transformations as a special
limiting case. In this way the EMUSY becomes able to connect various
possible clusterizations of different multicluster type, as well as
various many-particle configurations between the shell, collective
and cluster models of nuclear structure. The theory is briefly
illustrated using the nuclear system $^{24}$Mg as an example.
\end{abstract}
\maketitle

PACS number(s): {21.10.Re, 21.60.Gx, 21.60.Fw}

\section{Introduction}

The multichannel dynamical symmetry (MUSY)
\cite{musy-1,musy-2,musy-3a,musy-3b}, later termed
multiconfiguration dynamical symmetry \cite{musy-3b}, is a composite
symmetry that relates different clusterizations or fragmentations in
the many-fermion nuclear systems. The MUSY has been introduced at
the phenomenological level as a connecting symmetry of different
cluster configurations of the same type (particularly, two-cluster)
Ref.\cite{musy-1}. The channel in that reference refers to the
reaction channel, which defines a binary clusterization. It connects
the wave functions and the corresponding energy spectra of different
clusterizations in different energy windows in a unified way. In
Ref.\cite{musy-2} it was shown also that different binary
clusterizations can be related to each other in terms of a
subsidiary ternary configuration by projection.

The structure of MUSY consists of two important pieces
\cite{musy-1,musy-2,musy-3b}: i) an $U(3)$ dynamical symmetry for
each clusterization; and ii) a transformation in the particle-index
space, connecting the clusterizations under consideration. The
latter transformation for the particular case of binary
clusterizations is known in the literature as a Talmi-Moshinsky
transformation
\cite{Talmi-Moshinsky-1,Talmi-Moshinsky-2,Talmi-Moshinsky-3,Talmi-Moshinsky-4,musy-2}.
The multichannel (or multiconfiguration) dynamical symmetry appears
as a consequence of the antisymmetry of the nuclear wave function.
The MUSY thus transforms the wave function of one clusterization to
that of another clusterization. The next consequence of the
underlying $U(3)$ dynamical symmetry of MUSY is the generation of
identical spectra for different clusterizations or many-particle
nuclear configurations. Actually this argument was the
phenomenological basis for introducing the MUSY in 1994
\cite{musy-1}. The first ingredient of MUSY, i.e. the existence of
an $U(3)$ dynamical symmetry for each nuclear configuration, is
based on the fundamental role played by the three-dimensional
harmonic oscillator in nuclear physics.

Recently, a fully microscopic symplectic symmetry approach to
clustering (SSAC) was proposed for two- \cite{ssac} and general
multicluster \cite{multi-ssac} nuclear systems. The purpose of the
present work is to show that the SSAC allows to establish not only
the relationships between the shell, collective and cluster degrees
of freedom among each other in a natural way, but allows also to
determine a rigorous mathematical background of the extended
multiconfigurational dynamical symmetry (EMUSY). The latter, in
contrast to the original MUSY, allows to relate not only different
clusterizations of the same kind (e.g., two-cluster or binary), but
also connects various possible clusterizations of different
multicluster type. This allows to refer the present EMUSY as an
extended multiconfigurational dynamical symmetry, just as in Ref.
\cite{musy-3b}, instead of as a multichannel dynamical symmetry,
since in the limiting case of no clustering, one recovers the shell
and/or collective many-particle configurations (see, e.g., Ref.
\cite{multi-ssac}). Indeed, the MUSY represents a common
intersection of the shell, collective and cluster models of nuclear
structure \cite{musy-3a,musy-3b}. The same is valid for EMUSY too.

The EMUSY, likewise the MUSY, results from the antisymmetrization of
the many-nucleon wave functions. This was recognized by many authors
in the early days of the development of microscopic cluster theory
\cite{Perring56,Wildermuth58,Bayman58,Wildermuth62}. In particular,
it was even realized before \cite{Perring56} the introduction of
microscopic cluster model by K. Wildermuth and T. Kanellopoulos in
1958 that the cluster-model and the shell-model wave functions
become indistinguishable after the antisymmetrization. The
equivalence between the cluster and shell model wave functions has
further been established by means of the Bayman-Bohr theorem using
their $SU(3)$ symmetry character \cite{Bayman58}. The question of
how different models of nuclear structure are related to each other
was posed and realized early in the nuclear physics. In 1958 the
relation between the cluster and shell models was established by K.
Wildermuth and T. Kanellopoulos \cite{Wildermuth58} from one side,
and between the collective and shell models  by J. P. Elliott
\cite{Elliott58} from another. The question concerning the relation
between the various nuclear structure models has also been posed by
K. Wildermuth \cite{Wildermuth62}, and, in addition, using the
ground state of $^{16}$O as an example, it was asked how different
cluster configurations are related to each other, which is in some
sense the idea of MUSY (proto-MUSY) introduced later in depth by J.
Cseh \cite{musy-1,musy-2,musy-3a}. K. Wildermuth in
\cite{Wildermuth62} showed that due to the antisymmetrization the
wave functions between different clusterizations become identical,
as well as that the same is true for the shell-model and
cluster-model wave functions. The reason for this is the
indistinguishability of nucleons, i.e. the Pauli principle.
Recently, the idea of MUSY was implicitly considered
\cite{Horiuchi14} and exploited
\cite{Berriel-Aguayo21,ssac,multi-ssac} in the construction of the
many-particle Pauli allowed subspaces of the nuclear Hilbert space,
using the $SU(3)$ matching condition \cite{SACMa,SACMb} with the
shell model or equivalent to it procedure within the multicluster
SSAC for the case of no clustering \cite{multi-ssac}.

In the present work it will be shown also that the number-preserving
(unitary) MUSY transformations of Refs. \cite{musy-2,musy-3b},
connecting different nuclear configurations, are a particular case
of the more general symplectic transformations of EMUSY that appear
within the SSAC. The new theory is further illustrated using the
nuclear system $^{24}$Mg as an example. It will be demonstrated that
the symplectic transformations, connecting different many-particle
nuclear configurations, can be expressed in terms of the generators
in the enveloping algebra of the subgroup $Sp(6,R)_{R} \otimes
Sp(6,R)_{C} \subset Sp(6(A-1),R)$ only, where $Sp(6(A-1),R)$ is the
dynamical group of the whole many-nucleon nuclear system. This
simplifies the treatment of EMUSY and provides a more clear physical
interpretation.

\section{EMUSY within the SSAC}

We consider the two substructures of MUSY within the framework of
the symplectic symmetry approach to clustering in atomic nuclei for
the general case of nuclear system consisting of $k$ clusters.

\subsection{Dynamical symmetry}

In order to avoid the problem of the center-of-mass motion the
$m=A-1$ translationally invariant relative Jacobi coordinates
$q_{is}$ of the whole nuclear system are considered. Then, $(k-1)$
of the $(A-1)$ Jacobi vectors $\{\textbf{q}\} \equiv
\{\textbf{q}_{1}, \textbf{q}_{2}, \ldots, \textbf{q}_{A-1}\}$,
denote them by the set $\{\textbf{q}^{R}\}=
\{\textbf{q}^{R}_{1},\textbf{q}^{R}_{2},\ldots,\textbf{q}^{R}_{k-1}\}$,
will describe the relative motion of the $k$ clusters. The rest
$(A-k)$ Jacobi vectors will be related to the internal states of the
clusters \cite{multi-ssac}. The SSAC for the case of $k$-cluster
nuclear systems, i.e. when $A=A_{1}+A_{2} +\ldots +A_{k}$, can be
obtained by reducing the $Sp(6(A-1),R)$ dynamical group of the whole
nuclear system in the following way:
\begin{align}
&Sp(6(A-1),R) \notag\\
\notag\\
&\supset U(3(A-1)) \supset U_{R}(3(k-1)) \otimes U_{C}(3(A-k)) \notag\\
\notag\\
&\supset \quad \quad U_{R}(3) \quad \otimes \quad U_{C}(3) \notag\\
&\quad [E^{R}_{1},E^{R}_{2},E^{R}_{3}] \quad [E^{C}_{1},E^{C}_{2},E^{C}_{3}] \notag\\
\notag\\
&\supset \ \ U(3) \qquad\quad \supset \qquad SO(3).  \label{Sp6mR-U3m-RU3km1xCU3Amk}\\
&\ [E_{1},E_{2},E_{3}] \quad \ \ \kappa \qquad\quad \ L \notag
\end{align}
Eq. (\ref{Sp6mR-U3m-RU3km1xCU3Amk}) defines the $U(3)$ dynamical
symmetry as a first ingredient of the EMUSY. We note that the
permissible irreps of $U_{R}(3)$ for $k \geq 4$ are generally
three-rowed, whereas for $k = 3$ and $k = 2$ the permissible
$U_{R}(3)$ irreps are correspondingly two- and one-rowed. Similarly,
for clusters consisting of more than or equal to four nucleons, the
$U_{C}(3)$ permissible irreps are generally three-rowed and are
determined by the Pauli principle. The reduction chain
(\ref{Sp6mR-U3m-RU3km1xCU3Amk}) differs at first sight from that of
Ref. \cite{multi-ssac}, in which the multicluster framework within
the SSAC has been proposed. As will see, the present reduction chain
defined by Eq.(\ref{Sp6mR-U3m-RU3km1xCU3Amk}) turns out to be more
convenient for revealing the mathematical structure of the EMUSY,
compared to that of MUSY \cite{musy-1,musy-2,musy-3a,musy-3b}.

The $Sp(6(A-1),R)$ spectrum-generating algebra (SGA) is defined by
the following set of generators
\begin{align}
&F_{is,jt} = b^{\dag}_{is}b^{\dag}_{jt}, \label{Sp6mR-Fs} \\
&G_{is,jt} = b_{is}b_{jt}, \label{Sp6mR-Gs} \\
&A_{is,jt} = \frac{1}{2}\big(b^{\dag}_{is}b_{jt} +
b_{jt}b^{\dag}_{is}\big) \label{Sp6mR-As}
\end{align}
in terms of the standard creation $b^{\dag}_{is}=
\sqrt{\frac{M\omega}{2\hbar}} \big(q_{is}-\frac{i}{M\omega}p_{is}
\big)$ and annihilation $b^{\dag}_{is}=
\sqrt{\frac{M\omega}{2\hbar}} \big(q_{is}-\frac{i}{M\omega}p_{is}
\big)$ operators of harmonic oscillator quanta, where $i,j=1,2,3$
and $s,t=1,2,\ldots,m=A-1$. The operators $F_{is,jt}$ and
$G_{is,jt}$ create and annihilate a par of oscillator quanta. The
operators $A_{is,jt}$ annihilate a quantum at the $jt$ and create it
at $is$ degrees of freedom, i.e. shift a quantum from $jt$ to $is$
position. We can relabel the set $\{\textbf{q}\} \equiv
\{\textbf{q}_{1}, \textbf{q}_{2}, \ldots, \textbf{q}_{A-1}\}$ of
$3(A-1)$ relative Jacobi many-particle coordinates of nuclear system
as $\{\textbf{q}^{R}_{1},
\textbf{q}^{R}_{2},\ldots,\textbf{q}^{R}_{k-1},\textbf{q}^{C}_{1},
\textbf{q}^{C}_{2},\ldots,\textbf{q}^{C}_{A-k}\}$, corresponding to
the separation of the many-particle variables to dynamical
(collective) and kinematical (internal) ones, i.e. $\{\textbf{q}\} =
\{\textbf{q}_{D},\textbf{q}_{K}\}$, along the chain
(\ref{Sp6mR-U3m-RU3km1xCU3Amk}). It is useful to use alternative
labeling for the indices: $\xi,\xi'\equiv s,t=1,2,\ldots, k-1$ and
$s',t'\equiv s,t = k, k+1, \ldots, A-1$ or $1, 2, \ldots, A-k$
ranging over the $R$- and $C$-subsystem, respectively. Then, for
instance, the operators $A_{is,jt} \equiv A_{i\xi,jt'} =
b^{\dag}_{i\xi}b_{jt'}$ shift a quantum from $C$- to $R$-subsystem,
i.e. from $q^{C}_{jt'}$ to $q^{R}_{i\xi}$ coordinate.

\subsection{Transformations between different configurations}

The second component of the MUSY is the transformation, which
connects different clusterizations of the same type. These
transformations are represented simply by symplectic symmetry
transformations in the particle-index space within the SSAC, lying
in the enveloping algebra of the corresponding symplectic dynamical
group. To obtain the relevant transformation structure within the
SSAC we embed the substructure $U(3(A-1)) \supset U_{R}(3(k-1))
\otimes U_{C}(3(A-k))$ of Eq.(\ref{Sp6mR-U3m-RU3km1xCU3Amk}) in
$Sp(6(A-1),R) \supset Sp(6(k-1),R)_{R} \otimes Sp(6(A-k),R)_{C}$ and
consider the following reduction chain:
\begin{align}
&Sp(6(A-1),R) \notag\\
\notag\\
&\supset Sp(6(k-1),R)_{R} \otimes Sp(6(A-k),R)_{C} \notag\\
\notag\\
&\supset U_{R}(3(k-1)) \otimes U_{C}(3(A-k)) \notag\\
\notag\\
&\supset \quad \quad U_{R}(3) \quad \otimes \quad U_{C}(3) \notag\\
&\quad [E^{R}_{1},E^{R}_{2},E^{R}_{3}] \quad [E^{C}_{1},E^{C}_{2},E^{C}_{3}] \notag\\
\notag\\
&\supset \ \ U(3) \qquad\quad \supset \qquad SO(3) . \label{Sp6mR-RSp6km1xCSp6Amk}\\
&\ [E_{1},E_{2},E_{3}] \quad \ \ \kappa \qquad\quad \ L \notag
\end{align}
The equations (\ref{Sp6mR-U3m-RU3km1xCU3Amk}) and
(\ref{Sp6mR-RSp6km1xCSp6Amk}) differ from each other by the segments
$U(3(A-1)) \supset U_{R}(3(k-1)) \otimes U_{C}(3(A-k))$ and
$Sp(6(k-1),R)_{R} \otimes Sp(6(A-k),R)_{C} \supset U_{R}(3(k-1))
\otimes U_{C}(3(A-k))$. The reductions defined by these two
equations are actually equivalent, as can be seen from the following
common intersection:
\begin{align}
&Sp(6(A-1),R) \supset Sp(6(k-1),R)_{R} \otimes Sp(6(A-k),R)_{C}  \notag\\
&\qquad\quad \cup \qquad\qquad\qquad\quad \cup \qquad\qquad\qquad\qquad \cup \notag\\
&\quad U(3(A-1)) \ \ \supset U_{R}(3(k-1)) \quad \ \ \otimes \quad
U_{C}(3(A-k)). \label{intersection}
\end{align}
The $Sp(6(k-1),R)_{R}$ SGA is defined by the following set of
generators
\begin{align}
&F^{R}_{i\xi,j\xi'} = b^{\dagger R}_{i\xi}b^{\dagger
R}_{j\xi'}, \label{R-Sp6km1R-Fs} \\
&G^{R}_{i\xi,j\xi'} = b^{R}_{i\xi}b^{R}_{j\xi'}, \label{R-Sp6km1R-Gs} \\
&A^{R}_{i\xi,j\xi'} = \frac{1}{2}\big(b^{\dagger
R}_{i\xi}b^{R}_{j\xi'} +b^{R}_{j\xi'}b^{\dagger R}_{i\xi}\big),
\label{R-Sp6km1R-As}
\end{align}
where $i,j=1,2,3$ and $\xi,\xi'=1,2,\ldots,k-1$. When a summation
over the spatial or particle index is performed, one obtains the
groups $Sp(6,R)_{R} \equiv \{F^{R}_{ij} = \sum_{\xi}
F^{R}_{i\xi,j\xi}, G^{R}_{ij} = \sum_{\xi} G^{R}_{i\xi,j\xi},
A^{R}_{ij} = \sum_{\xi} A^{R}_{i\xi,j\xi}\}$ and $Sp(2(k-1),R)_{R}
\equiv \{F^{R}_{\xi\xi'} = \sum_{i}F^{R}_{i\xi,i\xi'},
G^{R}_{\xi\xi'} = \sum_{i}G^{R}_{i\xi,i\xi'}, A^{R}_{\xi\xi'} =
\sum_{i}A^{R}_{i\xi,i\xi'}\}$, respectively. The frequencies
$\omega_{R_{\xi}}$, entering in the definitions of $b^{\dagger
R}_{i\xi}=
\sqrt{\frac{\mu_{\xi}\omega_{R_{\xi}}}{2\hbar}}\Big(q^{R}_{i\xi}
-\frac{i}{\mu_{\xi}\omega_{R_{\xi}}}p^{R}_{i\xi}\Big)$ and
$b^{R}_{j\xi'} = (b^{\dagger R}_{j\xi'})^{\dagger}$, are chosen so
that for the oscillator length parameter one obtains $b_{0} =
\sqrt{\frac{\hbar}{\mu_{\xi}\omega_{R_{\xi}}}}=
\sqrt{\frac{\hbar}{M\omega}}$ for all $\xi = 1,\ldots,k-1$
\cite{multi-ssac}.

The group $Sp(6(A-k),R)_{C}$ is a direct-product group, i.e.
$Sp(6(A-k),R)_{C} = Sp(6(A_{1}-1),R)_{C_{1}} \otimes
Sp(6(A_{2}-1),R)_{C_{2}} \otimes \ldots \otimes
Sp(6(A_{k}-1),R)_{C_{k}}$. Each $Sp(6(A_{\alpha}-1),R)_{C_{\alpha}}$
SGA is defined by the set
\begin{align}
&F^{C_{\alpha}}_{is',jt'} =
b^{\dagger}_{is'}b^{\dagger}_{jt'}, \label{C-Sp6R-Fs} \\
&G^{C_{\alpha}}_{is',jt'} = b_{is'}b_{jt'},  \label{C-Sp6R-Gs} \\
&A^{C_{\alpha}}_{is',jt'} = \frac{1}{2}
\big(b^{\dagger}_{is'}b_{jt'} +b_{jt'}b^{\dagger}_{is'}\big) ,
\label{C-Sp6R-As}
\end{align}
where $s',t'= 1, 2, \ldots, A_{\alpha}-1$ and $\alpha = 1, 2,
\ldots, k$. Similarly, by contraction with respect to the spatial or
particle index, one obtains the groups $Sp(6,R)_{C_{\alpha}} \equiv
\{F^{C_{\alpha}}_{ij} = \sum_{s'}F^{C_{\alpha}}_{is',js'},
G^{C_{\alpha}}_{ij} = \sum_{s'}G^{C_{\alpha}}_{is',js'},
A^{C_{\alpha}}_{ij} = \sum_{s'}A^{C_{\alpha}}_{is',js'}\}$ and
$Sp(2(A_{\alpha}-1),R)_{C_{\alpha}} \equiv \{F^{C_{\alpha}}_{s't'} =
\sum_{i} F^{C_{\alpha}}_{is',it'}, G^{C_{\alpha}}_{s't'} = \sum_{i}
G^{C_{\alpha}}_{is',it'}, A^{C_{\alpha}}_{s't'} = \sum_{i}
A^{C_{\alpha}}_{is',it'}\}$, respectively. One can write down for
the whole $C$-subsystem $Sp(2(A-k),R)_{C} \equiv \{F^{C}_{s't'} =
\sum_{i} F^{C}_{is',it'}, G^{C}_{s't'} = \sum_{i} G^{C}_{is',it'},
A^{C}_{s't'} = \sum_{i} A^{C}_{is',it'}\}$, where $s',t'= 1, 2,
\ldots, A-k$ instead of $s',t'= 1, 2, \ldots, A_{\alpha}-1$.

Now observe that the $U_{R}(3(k-1)) \supset U_{R}(3) \otimes
U_{\xi}(k-1)$ substructure in the chain $U(3(A-1)) \supset
U_{R}(3(k-1)) \otimes U_{C}(3(A-k))$ of $Sp(6(A-1),R)$ coincides
with the reduction $U_{R}(6) \supset U_{R}(3) \otimes U_{p}(2)$ of
Ref. \cite{musy-2} for the case of $U_{C}(3)$ scalar clusters. The
ternary cluster configuration, connecting two different binary
clusterizations, was further defined by the structure $U_{R}(6)
\supset U_{R}(3) \otimes U_{p}(2)$ embedded in $U_{R}(7)$ and which
describes the two-relative motions of the three clusters of the
nuclear system \cite{musy-2}. The multicluster SSAC analog of the
reduction $U_{R}(6) \supset U_{R}(3) \otimes U_{p}(2)$ is thus
provided by $U_{R}(3(k-1)) \supset U_{R}(3) \otimes U_{\xi}(k-1)$,
where the group $U_{\xi}(k-1)$ coincides with the pseudospin
$U_{p}(2)$ group of Ref. \cite{musy-2} for the three-cluster ($k =
3$) configuration. Obviously, the groups $Sp(2(k-1),R)_{R}$ and
$Sp(2(A-k),R)_{C}$ extend the number-preserving unitary
transformations in the particle-index space, generated by the groups
$U_{\xi}(k-1)$ and $U(A-k)$, respectively. Note that the group
$Sp(2(k-1),R)_{R}$ changes the particle configurations within the
$R$-subsystem, whereas the group $Sp(2(A-k),R)_{C}$ alters the
particle configurations within the $C$-subsystem. Applying multiple
tensor products of symplectic generators of $Sp(2(k-1),R)_{R}$ and
$Sp(2(A-k),R)_{C}$ groups, which lie in the corresponding enveloping
algebras, one is able to obtain various clusterizations of arbitrary
type. For instance, in the particular case of $U(3)$-scalar clusters
and $k=3$, one obtains the extension of the $U_{R}(6) \supset
U_{R}(3) \otimes U_{p}(2)$ symmetry transformations in the
particle-index space of Refs. \cite{Talmi-Moshinsky-3,musy-2} to the
$Sp(12,R)_{R}$ symmetry transformations, which contain the
number-preserving $U_{R}(6)$ transformations as a subset. The
extension of the group structure $U_{R}(6) \supset U_{R}(3) \otimes
U_{p}(2)$ within the SSAC is then provided by the reductions
$Sp(12,R)_{R} \supset Sp(6,R)_{R} \otimes O_{p}(2)$ and
$Sp(12,R)_{R} \supset SO(3) \otimes Sp(4,R)_{p}$, where the
$Sp(4,R)_{p}$ transformations indicate that cluster configurations
with different type can be generated. The group $Sp(6,R)_{R}$
defines the spatial inter-cluster excitations, whereas the group
$Sp(4,R)_{p}$ changes the clusterization type. In the limiting case,
the group $Sp(4,R)_{p}$ contains as a subset the particular set of
number-preserving transformations, generated by its subgroup
$U_{p}(2) \subset Sp(4,R)_{p}$ and considered, e.g., in Refs.
\cite{Talmi-Moshinsky-3,musy-2}. The mutual relationships of
different subgroups in the three reduction chains $U_{R}(6) \supset
U_{R}(3) \otimes U_{p}(2)$, $Sp(12,R)_{R} \supset SO(3) \otimes
Sp(4,R)_{p}$, and $Sp(12,R)_{R} \supset Sp(6,R)_{R} \otimes
O_{p}(2)$ are established by the following lattice:
\begin{align}
&Sp(12,R)_{R} \ \supset SO(3) \otimes Sp(4,R)_{p} \notag\\
&\qquad \cup \qquad\qquad \cap \qquad\qquad \cup \notag\\
&\ \ U_{R}(6) \quad \ \ \supset \ U_{R}(3) \otimes \ U_{p}(2)
\notag\\
&\qquad \cap \qquad\qquad \cap \qquad\qquad \cup \notag\\
&Sp(12,R)_{R} \supset Sp(6,R)_{R} \otimes O_{p}(2).
\label{Sp12R-lattice}
\end{align}

But, instead of the reductions of Eqs.
(\ref{Sp6mR-U3m-RU3km1xCU3Amk}) and (\ref{Sp6mR-RSp6km1xCSp6Amk}),
it turns out that one can use an alternative reduction, which makes
a more transparent contact to the previous formulation of the SSAC
\cite{ssac,multi-ssac} and, as will see in the next section, is
simpler in many aspects. To make more clear this statement we recall
that the transformations in the particle-index space, connecting
different nuclear configurations or clusterizations within MUSY are
by construction $U(3)$-scalars, i.e. they preserve the underlying
$U(3)$ dynamical symmetry of the distinct configurations. Thus, the
Hamiltonian and transition operators expressed in terms of the
$U(3)$ SGA within different algebraic models will not be affected by
these connecting (particle-index space) transformations, and, hence,
will produce the same energy spectrum. The latter is crucial for
MUSY, resulting in physical operators that are invariant under the
transformations in particle-index space. This situations needs to be
considered in more detail. For simplicity, consider first the case
when there is no clustering, i.e. having a mononucleus within the
shell model. In the shell model different nucleon configurations
correspond to different clusterizations in the cluster models. For
$m$ fermions the unitary transformations that change the particle
configurations constitute the group $U(m)$. From other side, the
group of three-dimensional harmonic oscillator is $U(3)$, associated
with the possible distributions of oscillator quanta. It is known
that there is a close relationship between the distribution of
fermions and distribution of oscillator quanta, which for the
nuclear shell model is well demonstrated in Ref. \cite{UNtoSU3a}.
From the latter reference, it follows that the oscillator quanta are
distributed over the available shell-model levels in such a way as
if they were fermions (i.e., up to 2 or 4 on each level depending on
whether the proton-neutron or supermultiplet scheme of filling is
used).

The mathematical basis for relating the boson (oscillator quanta)
and fermion symmetries, i.e. between the groups $U(3)$ and $S_{A}$,
having a direct physical consequence for MUSY is carried out by
means of the $U(m)$ group by considering the shell-model coupling
scheme $U(3m) \supset U(3) \otimes U(m)$
\cite{Van71,Asherova75,Filippov81,GC,Van88,MS96-HObook}. The two
groups $U(3)$ and $U(m)$ are embedded in the larger unitary group
$U(3m)$, the symmetry group of a system of $m$ fermion particles in
the three-dimensional harmonic oscillator potential. When the
translationally-invariant Jacobi coordinates are used, i.e. when $m
= A-1$, we obtain the translationally-invariant shell model
\cite{Van71}. The underlying fermion symmetry of the nucleus is then
ensured by the reduction $U(m) \supset O(m) \supset S_{A}$ (or
simply $U(m) \supset S_{A}$). Because of the embedding $U(3) \otimes
U(m) \subset U(3m)$, the irreps of $U(3)$ are in a one-to-one
correspondence to the irreps of $U(m)$ and the same quantum numbers
are used to label the representations of both groups. The two groups
$U(3)$ and $U(m)$ are said to be mutually complimentary \cite{MQ70}
within the symmetric irreducible representations of $U(3m)$. The
Pauli principle within the shell model is fulfilled by providing a
minimum value of the number of oscillator quanta $N_{0}$ and all
permissible shell-model states have number of oscillator quanta $N
\geq N_{0}$, possessing at the same time the proper physical
permutational symmetry of the type
$[4^{k_{4}},3^{k_{3}},2^{k_{2}},1^{k_{1}}]$ ensured by the reduction
$U(m) \supset O(m) \supset S_{A}$.

In microscopic cluster models, the group $U(3m)$ is replaced by the
direct-product group $U(3m_{R}) \otimes U(3m_{C})$. More precisely,
we obtain the reduction  $U(3m) \supset U(3m_{R}) \otimes
U(3m_{C})$, when $m =m_{R} +m_{C}$. Setting $m_{R}=(k-1)$ and $m_{C}
= (A-k)$, we encounter the SSAC group structure $U_{R}(3(k-1))
\otimes U_{C}(3(A-k))$ of Eq.(\ref{Sp6mR-U3m-RU3km1xCU3Amk}). In the
limiting case of a scalar $U(3m_{R})$ part, we recover the shell
model.

The above considerations show that we can use not only the
transformations in the particle-index space, as stated in Refs.
\cite{musy-2,musy-3b}, but equivalently we can exploit the
corresponding transformations in the real three-dimensional space
too. This is so because the change of particle configuration results
in the change of the corresponding distribution of oscillator
quanta.

\vspace*{\fill}

\subsection{Alternative reduction}

We thus consider the following equivalent reduction chain
\begin{align}
&Sp(6(A-1),R) \notag\\
\notag\\
&\supset Sp(6(k-1),R)_{R} \otimes Sp(6(A-k),R)_{C} \notag\\
\notag\\
&\supset Sp(6,R)_{R} \otimes Sp(6,R)_{C} \otimes O(A-k)  \notag\\
\notag\\
&\supset \quad \quad U_{R}(3) \quad \otimes \quad U_{C}(3) \notag\\
&\quad [E^{R}_{1},E^{R}_{2},E^{R}_{3}] \quad [E^{C}_{1},E^{C}_{2},E^{C}_{3}] \notag\\
\notag\\
&\supset \ \ U(3) \qquad\quad \supset \qquad SO(3) , \label{Sp6mR-RSp6km1xCSp6Amk-RSp6RxCSp6RxOAmk}\\
&\ [E_{1},E_{2},E_{3}] \quad \ \ \kappa \qquad\quad \ L \notag
\end{align}
where the essential physics is represented by the group structure
$Sp(6,R)_{R} \otimes Sp(6,R)_{C} \otimes O(A-k)$. We note that all
three groups entering in the latter are direct-product groups:
$Sp(6,R)_{R} \equiv Sp(6,R)_{R_{1}} \otimes \ldots \otimes
Sp(6,R)_{R_{k-1}}$, $Sp(6,R)_{C} \equiv Sp(6,R)_{C_{1}} \otimes
\ldots \otimes Sp(6,R)_{C_{k}}$ and $O(A-k) \equiv O(A_{1}-1)
\otimes \ldots \otimes O(A_{k}-1)$. The combined cluster dynamics
within the SSAC in this way is described by the group structure
$Sp(6,R)_{R} \otimes Sp(6,R)_{C}$, where $Sp(6,R)_{R}$ describes the
intercluster excitations, whereas $Sp(6,R)_{C}$ is associated with
the internal excitations of the clusters. From another side, the
proper permutational symmetry within the present approach is ensured
via the orthogonal group $O(A-k) \equiv O(A_{1}-1) \otimes
O(A_{2}-1) \otimes \dots \otimes O(A_{k}-1)$ by considering the
reductions $O(A_{\alpha}-1) \supset S_{A_{\alpha}}$ with
$\alpha=1,2,\ldots,k$, given in detail in Ref. \cite{multi-ssac}.

The reduction chain (\ref{Sp6mR-RSp6km1xCSp6Amk-RSp6RxCSp6RxOAmk})
allows a simpler interpretation of the MUSY and its extension within
the SSAC. Similarly to Eqs.(\ref{Sp6mR-U3m-RU3km1xCU3Amk}) and
(\ref{Sp6mR-RSp6km1xCSp6Amk}), it defines the first component of
MUSY $-$ the $U(3)$ dynamical symmetry \cite{musy-1,musy-2,musy-3b}.
The transformations, connecting different clusterizations or nuclear
configurations, due to the arguments given in the end of the
preceding section can effectively be represented by the set of
symplectic transformations lying in the enveloping algebra of the
subgroup $Sp(6,R)_{R} \otimes Sp(6,R)_{C} \subset Sp(6(A-1),R)$
only. For example, the tensor operator $[G^{R}_{2} \times
F^{C}_{2}]^{l=0, m=0}_{(0,2) \ (2,0) \ (\lambda_{t},\mu_{t})}$
defined along the chain
(\ref{Sp6mR-RSp6km1xCSp6Amk-RSp6RxCSp6RxOAmk}) with the basis states
$|\Gamma;[E^{R}_{1},E^{R}_{2},E^{R}_{3}]_{3},[E^{C}_{1}, E^{C}_{2},
E^{C}_{3}]_{3};[E_{1}, E_{2}, E_{3}]_{3};\kappa (LS)J \rangle$ will
shift two oscillator quanta from $R$- to $C$-subsystem. This will
become more transparent in the practical application of EMUSY.

\section{Application}

As an example, we consider $^{24}$Mg. The relevant low-lying states
in first approximations are described by the leading $SU(3)$
multiplet $(8,4)$. The nucleus $^{24}$Mg admits different cluster
configurations: 1) $^{20}$Ne + $\alpha$; 2) $^{16}$O + $^{8}$Be; 3)
$^{12}$C + $\alpha$ + $\alpha$ + $\alpha$; 4) $^{12}$C + $^{12}$C;
5) $6\alpha$; etc. We recall, that the $SU(3)$ multiplet $(8,4)$
along the two-cluster channel $^{20}$Ne + $\alpha \rightarrow$
$^{24}$Mg is obtained by considering the outer product
$(\lambda_{R},\mu_{R}) \otimes (\lambda_{C},\mu_{C}) = (8,0) \otimes
(8,0)$ \cite{ssac} and using the $SU(3)$ matching condition
\cite{SACMa,SACMb} with the shell-model basis states. The $SU(3)$
connection has been shown to appear as a natural consequence in the
multicluster SSAC \cite{multi-ssac}. The relevant
$(\lambda_{R},\mu_{R}) \otimes (\lambda_{C},\mu_{C}) = (8,0) \otimes
(8,0)$ multiplet follows from the underlying shell-model (fermion)
$SU(3)$ structure of the clusters and the Wildermuth condition
\cite{Wildermuth77}. The latter requires the minimum of $E_{R} =
E^{R}_{1}+E^{R}_{2}+E^{R}_{3}= 8$ oscillator quanta for the
intercluster excitation, allowed by the Pauli principle. Similarly,
for other channels one obtains the relevant $E_{R}$ and
$(\lambda_{R},\mu_{R}) \otimes (\lambda_{C},\mu_{C})$ quantum
numbers which are given in Table \ref{Mg24-channels}. We point out
also that the number of available oscillator quanta can be
represented as $E=E_{R}+E_{C}$, which for different clusterizations
is distributed in various ways among the $R$- and $C$-subsystems,
respectively. This is considered in detail in Ref. \cite{multi-ssac}
for the case of a single $4\alpha$ cluster configuration of the
$^{16}$O nuclear system. For $^{24}$Mg the total number of
oscillator quanta, corresponding to the leading $SU(3)$ multiplet
$(8,4)$, is $E=28$.

\begin{table}[h!]
\caption{Relevant $E_{R}$ and $(\lambda_{R},\mu_{R}) \otimes
(\lambda_{C},\mu_{C})$ quantum numbers along some of the possible
cluster channels for $^{24}$Mg.} \label{Mg24-channels}
\smallskip\centering\small\addtolength{\tabcolsep}{5.pt}
\begin{tabular}{l|l|l}
\hline\hline \qquad channel $ $ & $E_{R}$ & $(\lambda_{R},\mu_{R})
\otimes (\lambda_{C},\mu_{C})$
\\ \hline\hline
$^{20}Ne + \alpha \rightarrow ^{24}Mg$ & $8 $ & $\qquad (8,0)
\otimes (8,0)$
\\ \hline
$^{16}O + ^{8}Be \rightarrow ^{24}Mg$ & $12$ & $%
\begin{tabular}{l}
$\quad (12,0) \otimes (4,0)$%
\end{tabular}%
$ \\ \hline
$^{12}C + \alpha + \alpha + \alpha \rightarrow ^{24}Mg$ & $20$ & $%
\begin{tabular}{l}
$\quad (20,0) \otimes  (0,4)$%
\end{tabular}%
$ \\ \hline
$^{12}C + ^{12}C \rightarrow ^{24}Mg$ & $12$ & $%
\begin{tabular}{l}
$\quad (12,0) \otimes  (0,8)$%
\end{tabular}%
$ \\ \hline $6\alpha \rightarrow ^{24}Mg$ & $28$ & $\qquad (8,4)
\otimes  (0,0)$ \\ \hline
\end{tabular}%
\end{table}

Let introduce the following notations for the symplectic generators
$G^{R}_{2M} \equiv G^{R \qquad\quad \ \ 2M}_{(0,2) \ (0,0) \ (0,2)}$
and $F^{C}_{2M} \equiv F^{C \qquad\quad \ \ 2M}_{(0,0) \ (2,0) \
(2,0)}$ of the $Sp(6,R)_{R} \otimes Sp(6,R)_{C}$ SGA, which are
written as tensor operators with respect to different subgroups
along the chain (\ref{Sp6mR-RSp6km1xCSp6Amk-RSp6RxCSp6RxOAmk}). The
basis states can shortly be written down as
$|\Gamma;(\lambda_{R},\mu_{R}),(\lambda_{C},\mu_{C});(\lambda,\mu);\kappa
(LS)J \rangle$ \cite{multi-ssac} using the Elliott's $SU(3)$
notation. Other multiple tensor operators can be similarly written.
Now, for instance, the cluster configuration $^{20}$Ne + $\alpha$
with the multiplet structure $(8,0) \otimes (8,0)$ can be obtained
from the channel $^{16}$O + $^{8}$Be having the $(12,0) \otimes
(4,0)$ multiplet structure by acting with the tensor operator
$[[G^{R}_{2} \times G^{R}_{2}] \times [F^{C}_{2} \times
F^{C}_{2}]]^{l=0, m=0}_{(0,4) \ (4,0) \ (\lambda_{t},\mu_{t})}$ on
the latter configuration. The $SU(3)$ tensor properties
$(\lambda_{t},\mu_{t})$ of the coupled tensor operator in this case
can take the values $(\lambda_{t},\mu_{t}) = (4,4),(2,2),(0,0)$.
But, in order to fit more naturally to the MUSY concept, in which
the underlying $SU(3)$ dynamical symmetry is preserved, we chose
$(\lambda_{t},\mu_{t}) = (0,0)$ for this particular case. Thus, the
tensor operator $[[G^{R}_{2} \times G^{R}_{2}] \times [F^{C}_{2}
\times F^{C}_{2}]]^{l=0, m=0}_{(0,4) \ (4,0) \ (0,0)}$, preserving
the total number of oscillator quanta $E=28$, shifts 4 quanta from
$R$- to $C$-subsystem, but keeping unchanged the underlying $SU(3)$
dynamical symmetry represented by the leading $SU(3)$ cluster
multiplet $(8,4)$. Similarly, the two-cluster configuration $^{12}$C
+ $^{12}$C with the multiplet structure $(12,0) \otimes (0,8)$ can
be obtained from the clusterization $^{12}$C + $\alpha$ + $\alpha$ +
$\alpha$ having a $(20,0) \otimes (0,4)$ tensor character by means
of the tensor operator $[[G^{R}_{2} \times G^{R}_{2} \times
G^{R}_{2} \times G^{R}_{2}] \times [F^{C}_{2} \times F^{C}_{2}
\times F^{C}_{2} \times F^{C}_{2}]]^{l=0, m=0}_{(0,8) \ (0,4) \
(4,4)}$ that shifts 8 oscillator quanta from $R$- to $C$-subsystem.
Likewise, the $^{16}$O + $^{8}$Be configuration with the multiplet
structure $(12,0) \otimes (4,0)$ can be obtained from the
clusterization $^{12}$C + $^{12}$C having $(12,0) \otimes (0,8)$ by
acting on the latter by the tensor operator $[[G^{C}_{2} \times
G^{C}_{2} \times G^{C}_{2} \times G^{C}_{2}] \times [F^{C}_{2}
\times F^{C}_{2} \times F^{C}_{2} \times F^{C}_{2}]]^{l=0,
m=0}_{(0,0) \ (4,4) \ (4,4)}$ which redistributes the oscillator
quanta only within the $C$-subsystem and preserves $E_{R} = 12$.
Further, the clusterization $^{12}$C + $\alpha$ + $\alpha$ +
$\alpha$ with the multiplet structure $(20,0) \otimes (0,4)$ can be
obtained from the $6\alpha$ cluster configuration having $(8,4)
\otimes (0,0)$ by means of the more complicated tensor operator
$[[A^{R}_{2} \times A^{R}_{2} \times A^{R}_{2} \times A^{R}_{2}
\times G^{R}_{2} \times G^{R}_{2} \times G^{R}_{2} \times G^{R}_{2}]
\times [F^{C}_{2} \times F^{C}_{2} \times F^{C}_{2} \times
F^{C}_{2}]]^{l=0, m=0}_{(8,4) \ (0,4) \ (4,4)}$ which shifts 8
oscillator quanta from $R$- to $C$-subsystem and, in addition,
redistributes the remaining 20 quanta within the $R$-subsystem to
obtain the resulting $SU_{R}(3)$ multiplet $(20,0)$. Note that for
the $6\alpha$ configuration of $^{24}$Mg, consisting of scalar
$SU_{C}(3)$ clusters, the corresponding cluster dynamics reduces to
the intercluster dynamics of the $R$-subsystem only. Other
transformations between the various clusterizations of $^{24}$Mg can
be obtained in a similar fashion.

Note that all tensor operators consisting of the multilinear
products of the raising, lowering and number-preserving symplectic
generators of the $R$- and/or $C$-subsystem belong to the enveloping
algebra of the group $Sp(6,R)_{R} \otimes Sp(6,R)_{C}$ only. The
limiting cases are represented by the tensor operators acting only
within the one subsystem, i.e. $R$- or $C$-subsystem. In this
regard, we note that when no clusterization is assumed, $^{24}$Mg is
described by the $Sp(6,R)$ collective model \cite{RR1},
corresponding to the limiting case of SSAC with a scalar
$Sp(6,R)_{R}$ part. The relevant $(\lambda_{R},\mu_{R}) \otimes
(\lambda_{C},\mu_{C})$ multiplet structure in this case is $(0,0)
\otimes  (8,4)$. Then, the $Sp(6,R)_{R} \otimes Sp(6,R)_{C}$ group
structure, governing the nuclear dynamics, shrinks to $Sp(6,R)$ with
the natural identification $Sp(6,R) \equiv Sp(6,R)_{C}$
\cite{multi-ssac}. The nuclear dynamics therefore reduces to the
dynamics of the $C$-subsystem (mononucleus), i.e. the cluster
dynamics becomes a pure collective-model dynamics.

Similarly, one can obtain the other Pauli allowed $SU(3)$
multiplets, which as a result of the antisymmetrization become
identical for different clusterizations from one side, and for the
shell, cluster and collective model configurations from another
side. The present theory has demonstrated how the connecting
transformations of EMUSY within the SSAC can be obtained explicitly
for arbitrary types of clusterization of the multicluster nuclear
systems.

Finally, we stress that a crucial condition for the equivalence of
the many-particle wave functions in various nuclear structure models
is to use the same intercluster and intrinsic cluster oscillator
length parameters. Otherwise, the equivalence will be lost. The
latter is easily understood within the symplectic algebraic
framework, since the wave functions with different from the standard
shell-model oscillator length parameter $b_{0} =
\sqrt{\frac{\hbar}{M\omega}}$ are simply deformed wave functions
obtained from those of the spherical shell model by applying
symplectic scale transformations.

\section{Conclusions}

In the present paper an extended multiconfigurational dynamical
symmetry is introduced within the SSAC. EMUSY generalizes the MUSY
of Refs. \cite{musy-1,musy-2,musy-3a,musy-3b} in two aspects: 1) it
is valid for general multicluster nuclear systems; 2) it involves
more general symplectic transformations, connecting different
many-particle configurations in the shell, collective and cluster
models of nuclear structure. The EMUSY, likewise the MUSY, results
from the antisymmetrization of the many-nucleon wave functions.

A rigorous mathematical foundation of EMUSY within the SSAC was
given and compared to the mathematical structures of original MUSY.
It was shown that the symplectic transformations, connecting
different many-particle configurations, contain the standard
number-preserving (unitary) MUSY transformations as a special
limiting case. Further, the close relationship between the
symmetries in the three-dimensional and particle-index spaces was
considered in more detail. It was demonstrated also that because of
the mutual complementarity of the ordinary space and multiparticle
space symmetries, instead of the $Sp(2(k-1),R)_{R}$ and
$Sp(2(A-k),R)_{C}$ connecting transformations one can equivalently
use the symplectic $Sp(6,R)_{R} \otimes Sp(6,R)_{C}$
transformations. The latter significantly simplifies the physical
interpretation of EMUSY. This situation is similar to that of the
proton-neutron $SU_{p}(3) \otimes SU_{n}(3)$ shell model
\cite{pseudoSU3c}. If we take its symplectic extension, represented
by the $Sp(6,R)_{p} \otimes Sp(6,R)_{n}$ shell model, then the
symplectic raising and lowering generators of the latter will change
the $SU_{p}(3) \otimes SU_{n}(3)$ multiplet structure
$(\lambda_{p},\mu_{p}) \otimes (\lambda_{n},\mu_{n})$, corresponding
to the change of the underlying proton-neutron shell structure.
Resembling the case of $Sp(6,R)_{p} \otimes Sp(6,R)_{n}$, the
raising and lowering $Sp(6,R)_{R} \otimes Sp(6,R)_{C}$ generators
change the $SU_{R}(3) \otimes SU_{C}(3)$ multiplet structure
$(\lambda_{R},\mu_{R}) \otimes (\lambda_{C},\mu_{C})$ within the
SSAC, corresponding to the change of the underlying many-particle
cluster configuration. Note that the unitary $SU_{p}(3) \otimes
SU_{n}(3)$ and $SU_{R}(3) \otimes SU_{C}(3)$ transformations can not
change the $SU_{p}(3) \otimes SU_{n}(3)$ and $SU_{R}(3) \otimes
SU_{C}(3)$ multiplet structure, respectively.

The new theory was briefly illustrated using the nuclear system
$^{24}$Mg as an example. For this nucleus, it has been shown how the
connecting transformations between any two clusterizations of an
arbitrary type can be obtained explicitly. It has been demonstrated
also that some quantum numbers are not respected by the EMUSY, i.e.
the same quantum numbers are not valid at the same time in different
many-particle configurations, reflecting the symplectic nature of
the connecting transformations. In SSAC these quantum numbers are
generally represented by the separate numbers of oscillator quanta
for the $R$- and $C$-subsystems $E_{R}$ and $E_{C}$, respectively,
resulting in the change of the $SU_{R}(3) \otimes SU_{C}(3)$
multiplet structure $(\lambda_{R},\mu_{R}) \otimes
(\lambda_{C},\mu_{C})$. In this respect, we note that in the
limiting cases of $SU_{R}(3)$ or $SU_{C}(3)$ scalar parts, the
cluster dynamics within the SSAC reduces to the pure $C$- or
$R$-dynamics only. In the first case, the cluster dynamics becomes a
pure collective-model dynamics.

We point out that when the $U(3)$ or $SU(3)$ symmetry in
(\ref{Sp6mR-RSp6km1xCSp6Amk-RSp6RxCSp6RxOAmk}) is broken, depending
on the mixing interactions used, the produced energy spectrum could
be different for different nuclear configurations. But, even in this
case of broken $SU(3)$ symmetry, due to the antisymmetrization, the
EMUSY provides a connection between the basis states of different
nuclear configurations possessing the same $SU(3)$ character. So,
the many-particle subspaces of the nuclear Hilbert space will
consist of the same $SU(3)$ basis states. The connection between
these basis states for different clusterizations or nuclear
configurations (in the shell and collective models) generally will
require the usage of different nuclear interactions in the model
Hamiltonian. For instance, the $0 \hbar\omega$ many-particle
subspace of the Hilbert space of $^{24}$Mg for the $^{20}$Ne +
$\alpha$ channel contains also the $SU(3)$ multiplets $(4,6)$ and
$(0,8)$ \cite{ssac}, in addition to the leading one $(8,4)$. These
three $SU(3)$ representations can be mixed (connected) by
introducing a simple horizontal-mixing interaction $H_{hmix} =
h([G^{R}_{2}\times F^{R}_{2}]^{l=0, m=0}_{(2,2) \ (0,0) \ (2,2)} +
h.c.)$, expressed in terms of the $Sp(6,R)_{R}$ generators only.
This interaction, however, can not connect different $Sp(6,R)_{R}$
representations. Thus, to reach the $SU(3)$ multiplets $(4,6)$ and
$(0,8)$ from $(8,4)$ in the $6\alpha$ channel, from another side,
one needs to consider symplectic symmetry breaking interactions
(e.g., pairing) connecting different $Sp(6,R)_{R}$ irreducible
representations which bandheads are these three $SU(3)$ multiplets.
In addition, note also that the mixing interaction $H_{hmix}$ is
configuration dependent, hence it will produce a different energy
contribution for various clusterizations (different from $6\alpha$)
and their spectra will not be identical.

We hope that the present study will deepen our understanding of the
unifying ability of EMUSY and its physical interpretation.

\end{document}